 \documentclass{ws-ijmpa}

\usepackage[super,compress]{cite}
\usepackage{graphicx}
\begin{document}
\markboth{Vindhyawasini Prasad}{Highlights of recent BESIII results}

%
\catchline{}{}{}{}{}
%

\title{Highlights of recent BESIII results 
}

\author{Vindhyawasini Prasad}

\address{(On behalf of the BESIII Collaboration) \\
Instituto de Alta Investigation \\
Universidad de Tarapaca \\
Casilla 7D, Arica, Chile \\
vindymishra@gmail.com}

\maketitle

\begin{history}
\received{Day Month Year}
\revised{Day Month Year}
\end{history}

\begin{abstract}
This report highlights the recent BESIII results related to the measurement of the $R$ value, radiative decays of $J/\psi$ and the polarization of $D^0 \to \omega \phi$ decay. BESIII, a symmetric electron-positron collider experiment, has collected a large data sample at several energy points in the tau-charm region, including the $J/\psi$ and $\psi(3770)$ resonances.  Based on the data samples collected at 14 center-of-mass energy ($\sqrt{s}$) points between  2.2323 and 3.671 GeV, BESIII has achieved a measurement of the $R$ value with an accuracy down to the level of $2.6\%$ for $\sqrt{s} < 3.1$ GeV and $3.0\%$ for $\sqrt{s} > 3.1$ GeV. BESIII has also observed the $X(2600)$ resonance in radiative $J/\psi$ decay and the polarization of $D^0 \to \omega \phi$ for the first time. The polarization of $D^0 \to \omega \phi$ saturates with completely transverse polarization, which is inconsistent with the existing theoretical models. These results are useful in testing the predictions of the Standard Model.

\end{abstract}

\ccode{PACS numbers:}


\section{Introduction}	
Quantum chromodynamics (QCD) delineates the theory of hadrons that bind quarks together via the strong force mediated by gluons. The hadron production at electron-positron collision occurs via the annihilation of electron-positron pairs into a virtual $\gamma$ or $Z^0$ boson~\cite{zhang}. The $R$ value is the lowest-order cross-section ratio of the inclusive hadronic process $e^+e^- \to {\rm hadrons}$ to the quantum electrodynamics (QED) process $e^+e^- \to \mu^+\mu^-$~\cite{rvalue}. It is an important input for testing the predictions of the Standard Model (SM)~\cite{amu, mz}. The $R$ value below 5 GeV  is used in the theoretical calculation of the anomalous magnetic moment~\cite{amu}. The $R$ value also appears in the calculation of  the QED running coupling constant evaluated at the $Z$ pole~\cite{mz}. This observable can provide another precision test of the SM and is essential for the electroweak precision program at future colliders~\cite{futureRvale}.

QCD is a non-Abelian theory of self-interaction of gluons that can form the quarkless states of glueballs. Lattice QCD theory predicts the scalar (pseudoscalar or tensor) glueballs within the mass range of $1-2$ ($2-3$) GeV/$c^2$~\cite{lqcd}. Such quarkless glueballs can be accessible via radiative  $J/\psi$ decays~\cite{bfradjps}. The branching fraction for glueball production  in radiative $J/\psi$ decays is predicted to be within the range of $10^{-2} - 10^{-3}$, depending upon the masses and spin parity of the glueballs~\cite{bfradjps}.  A light $CP$-odd Higgs boson, predicted by many models beyond the SM, such as Next-to-Minimal Supersymmetric Standard Model (NMSSM)~\cite{NMSSM1, NMSSM2, NMSSM3},  can also be produced via radiative decays of $J/\psi$~\cite{wilczek}. An experimental constraint on such a light Higgs boson scenario is desirable to test the SM~\cite{Lisanti}.

The long-distance contributions to $D^0 - \bar{D}^0$ mixing~\cite{longdist} can be evaluated using two body hadronic decays of $D^0$ mesons, such as $D^0$ meson decaying into two vector ($V$) mesons. The two vector mesons in $D^0 \to VV$ decays produce a longitudinal partial-wave amplitude ($H_0$), which is $CP$-even, and two transverse partial wave amplitudes ($H_{\pm}$), which are superpositions of $CP$-even and $CP$-odd states. The polarization of $D^0 \to VV$ is sensitive to the $V-A$ structure of electroweak interactions, spin correlation, and final state interactions of the involved mesons~\cite{weakint1, weakint2, weakint3}. A previous measurement shows that the decay $D^0 \to \bar{K}^{*0} \rho^0$ is transversely polarized~\cite{mark3}, and a precise measurement of $D^0 \to \rho^0 \rho^0$ reaction indicated that the decay is dominated by longitudinal polarization~\cite{focus}. These measurements are inconsistent with the naive factorization~\cite{fact} and Lorentz invariant-based symmetry~\cite{LI} models, which predict the longitudinal polarization fraction, $f_L = H_0^2/(H_0^2 +H_-^2+H_+^2)$, to be $\sim 0.5$ and 0.33, respectively.

The large data samples collected by the BESIII experiment~\cite{bes3} with several energy points in the tau-charm region, including $J/\psi$ and $\psi(3770)$ resonances, provide an ideal avenue for studying and testing the precision of the SM.  This report highlights the recent results of the BESIII experiment related to the measurements of $R$ value, radiative decays of $J/\psi$ and polarization of $D^0 \to \omega \phi$.

\section{\boldmath{$R$} value measurement}
 The $R$ value  at low energies has been measured by many experiments~\cite{prevRvale}. The precision of the latest measurement of the $R$ value, measured by the KEDR experiment, is down to the level of $3.0\%$ above 3.1 GeV~\cite{kedr}. BESIII has recently measured the $R$ value using the data samples collected at 14 center-of-mass (CM) energy ($\sqrt{s}$) points between [2.23, 3.67] GeV~\cite{bes3rvale}. The $R$ value is calculated using the following formula

\begin{equation}
R = \frac{N_{\rm had}^{\rm obd}-N_{\rm bkg}}{\sigma_{\mu\mu}^0 \mathcal{L}_{\rm int} \epsilon_{\rm trig} \epsilon_{\rm had} (1+ \delta)},
\end{equation}

\begin{figure}[b]
\centerline{\includegraphics[width=12.0cm]{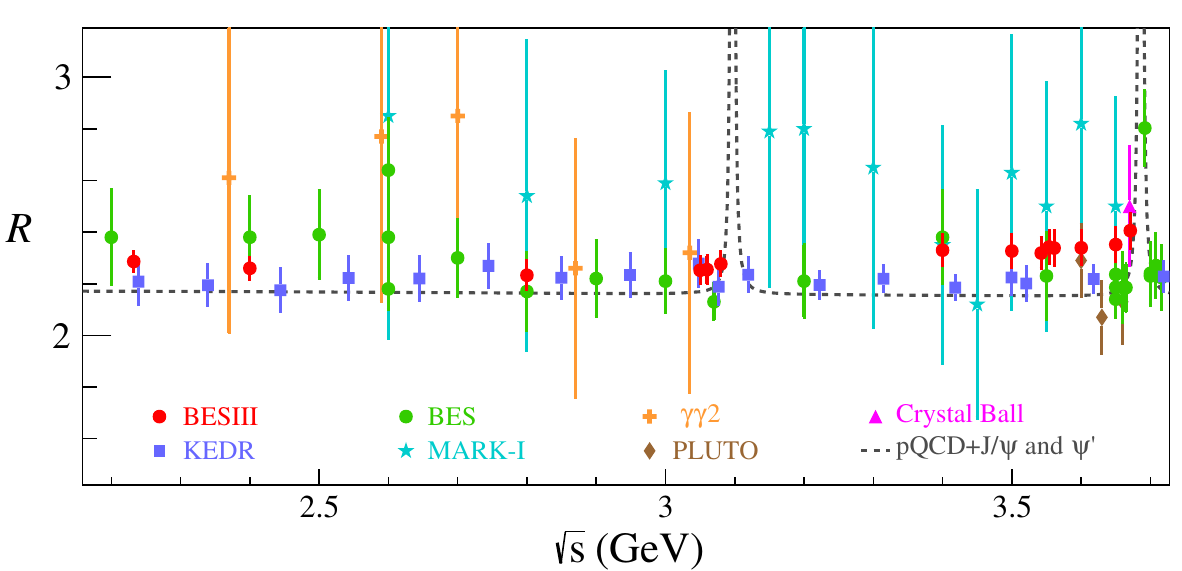}}
\caption{The measured $R$ value as a function of the center-of-mass energy between 2.2 and 3.7 GeV from various experiments, including the BESIII measurement~\cite{bes3rvale}. The $R$ value of the BESIII measurement is consistent with KEDR measurement~\cite{kedr} within the uncertainty below 3.1 GeV.}
\label{R_value}
\end{figure}

\noindent where $N_{\rm had}^{\rm obd}$ is the number of hadronic two- or more than two-prong events obtained directly from data, $N_{\rm bkg}$ is the remaining number of background events after applying all the selection criteria, $\sigma_{\mu\mu}^0 = 4\pi \alpha^2(0)/(3s)$ is the leading-order $e^+e^- \to \mu^+\mu^-$ cross section, $\mathcal{L}_{\rm int}$ is the integrated luminosity of the data sample, $\epsilon_{\rm trig}$ is the trigger efficiency for hadronic events, $\epsilon_{\rm had}$ is the detection efficiency for inclusive hadronic events, and $(1+\delta)$ is the initial state radiation (ISR) correction factor.  The hadronic detection efficiency measurement is based on two different simulation models, which are the existing inclusive {\sc LUARLW} generator~\cite{luarlw} and a new hybrid framework integrating a few exclusive generators. The detection efficiencies obtained from the two simulation models are consistent with each other down to the level of $2.3\%$.  The $R$ value measurement achieves an accuracy down to the level of $2.6\%$ below 3.1 GeV and $3\%$ above 3.1 GeV, as shown in Fig.~\ref{R_value}. The average $R$ value of the BESIII measurement in the CM energy range from 3.4 to 3.6 GeV is larger than the corresponding KEDR result~\cite{kedr} and theoretical expectation~\cite{Baikov} by 1.9 and 2.7 standard deviations, respectively. Improved precision of the measurement of the $R$ value is desirable to shed light on this discrepancy and improve the precision of the SM predictions of $\alpha(M_Z^2$)~\cite{mz} and anomalous magnetic moment~\cite{amu}. 

\section{Radiative \boldmath{$J/\psi$} decays}
The radiative decays of the $J/\psi$ meson are ideally suited for the studies of glueballs~\cite{bfradjps} and  new physics searches beyond the SM~\cite{NMSSM1, NMSSM2, NMSSM3}. Using 10 billion $J/\psi$ events, BESIII has recently observed several new glueball states and searched for a light Higgs boson. The following sub-sections describe the recent results of the BESIII experiment related to the radiative decays of $J/\psi$.

\begin{figure}[b]
\centerline{\includegraphics[width=12.0cm]{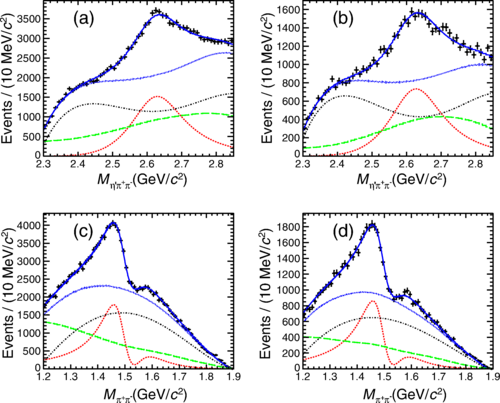}}
\caption{The simultaneous fit to the $\pi^+\pi^-\eta'$ and $\pi^+\pi^-$ spectra distributions with the two $\eta'$ decays modes of $\eta' \to \gamma \pi^+\pi^-$ (a, c) and $\eta' \to \pi^+\pi^- \eta$ (b, d).  The black dots with error bar are data, blue solid lines, the red dashed curves, black dash-dotted lines, green long dashed lines, and blue dotted lines correspond to the total fit result, $X(2600)$ signal, non-peaking backgrounds, $J/\psi \to \pi^0\pi^+\pi^-\eta'$ (or non-$\eta'$) backgrounds, and total backgrounds, respectively. }
\label{x2600}
\end{figure}

\subsection{Observation of the state \boldmath{$X(2600)$} in \boldmath{$J/\psi \to \gamma \pi^+\pi^- \eta'$}}
The $J/\psi \to \gamma \pi^+\pi^- \eta'$ decay is one of the golden modes to search for pseudoscalar glueballs~\cite{Amsler, Eshraim}. The exotic states  $X(1835)$, $X(2120)$ and $X(2370)$ were observed in $J/\psi \to \gamma \pi^+\pi^-\eta'$ decays by BESIII~\cite{bes3rad} and predecessor experiments~\cite{besrad}. The spin parity of the $X(1835)$ resonance was determined to be $0^{-+}$  in $J/\psi \to \gamma K_S^0 K_S^0 \eta$~\cite{spinx1835}, but its nature is still mystery.  Theoretical prediction of $X(1835)$ could be a $p\bar{p}$ bound state~\cite{pbound}, the second radial excitation of the $\eta'$~\cite{radexit}, or a pseudoscalar glueball~\cite{psedoglub}. The measured mass of the $X(2370)$ favours the lattice QCD prediction of the pseudoscalar glueball~\cite{lcqcd}. Further observations of the process $J/\psi \to \gamma \pi^+\pi^- \eta'$ are desirable to better understand the QCD and hadron physics.  Based on $(10087 \pm 44) \times 10^6$ $J/\psi$ events,  BESIII has recently studied the process $J/\psi \to \gamma \pi^+\pi^- \eta'$ with the two $\eta'$ decay modes: $\eta' \to \gamma \pi^+\pi^-$ and $\eta' \to \eta \pi^+\pi^-$, $\eta \to \gamma \gamma$~\cite{bes3x2600}. A new resonance, the  $X(2600)$, has been observed in the $\pi^+\pi^-\eta'$ invariant mass spectrum with a significance of more than $20\sigma$.  A simultaneous fit to the $\pi^+\pi^- \eta'$ and $\pi^+\pi^-$ invariant mass spectra with the two $\eta'$ decay modes indicate that the $X(2600)$ has a connection to a structure around 1.5 GeV/$c^2$ in the $\pi^+\pi^-$ invariant mass spectrum, as seen in Fig.~\ref{x2600}.  The structure around 1.5 GeV/$c^2$ in $\pi^+\pi^-$ invariant mass spectra is described with the interference between the $f_0(1500)$ and the $X(1540)$ resonances. The mass and width of the $X(2600)$ are determined to be $2618.3 \pm 2.0^{+16.3}_{-1.4}$ MeV/$c^2$ and $195 \pm 5^{+26}_{-17}$ MeV, where first and second uncertainties are statistical and systematic, respectively. Determining the spin parity of the $X(2600)$ would be crucial to understanding its exact nature.

\begin{figure}[b]
\centerline{\includegraphics[width=12.0cm]{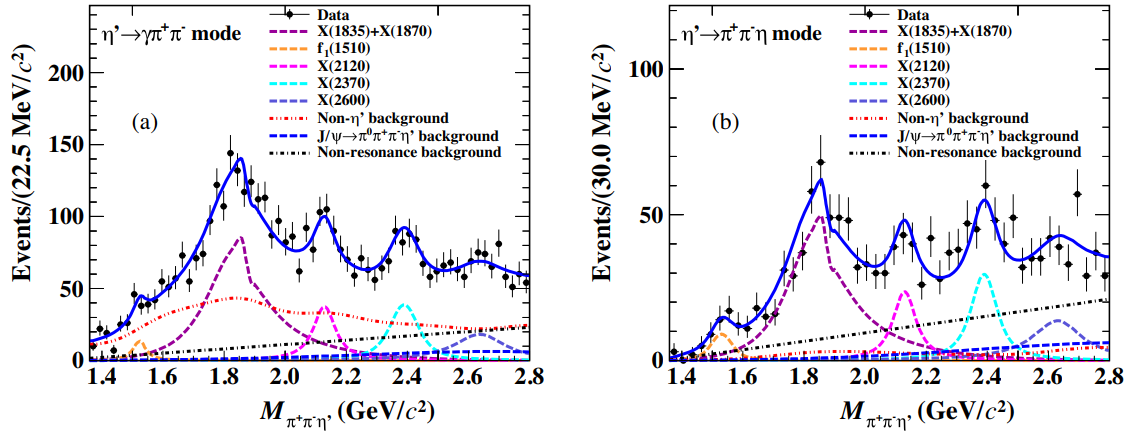}}
\caption{The simultaneous fit to the $\pi^+\pi^-\eta'$ invariant mass spectrum with the two $\eta'$ decays modes of $\eta' \to \gamma \pi^+\pi^-$ (left) and $\eta' \to \pi^+\pi^- \eta$ (right).  The black dots with error bar are data, the blue solid curve is the total fit result, the orange dashed curve represents $f_1(1510)$ resonance, the purple dashed line the $1835)$ and $X(1870)$ states, the pink dashed line the $X(2120)$ state, the cyan dashed line the $X(2370)$ resonance and the light blue dashed curve the $X(2600)$ state. The red dashed line, blue dashed line, and black dot-dashed lines correspond to the non-$\eta'$ background, $J/\psi \to \pi^0\pi^+\pi^-\eta'$ and other possible non-resonant backgrounds, respectively.  }
\label{dalitz}
\end{figure}

\subsection{$X(2600)$ resonance in the electromagnetic Dalitz decay of \boldmath{$J/\psi \to e^+e^- \pi^+\pi^- \eta'$}}
The existence of the $X(1835)$, $X(2120)$, $X(2370)$, and $X(2600)$ resonances is also confirmed by an alternative analysis of electromagnetic (EM) Dalitz decay of $J/\psi \to e^+e^- \pi^+\pi^- \eta'$ using 10 billion of $J/\psi$ events collected by the BESIII detector~\cite{EMDalitz}.  The electromagnetic Dalitz decays  $J/\psi \to e^+e^- \pi^+\pi^- \eta'$ proceed via $J/\psi \to \gamma^* \pi^+\pi^- \eta'$, where $\gamma^*$ is a virtual photon converting into electron-positron pair in the final state~\cite{Dalitzdecay}. The corresponding di-electron invariant mass spectrum dependent Transition Form Factor (TFF) deviates from the point-like prediction of the QED due to dynamics of the electromagnetic spectrum~\cite{Dalitzdecay}. Thus, the EM Dalitz decays are a sensitive probe of the inner structure of the mesons involved in the process.  In $J/\psi \to e^+e^- \pi^+\pi^- \eta'$ decay, the peaking background from $J/\psi \to \gamma \pi^+\pi^- \eta'$ decay mainly originates the beam pipe and inner wall of the main drift chamber of the BESIII detector. A gamma conversion finder algorithm is used to reject the backgrounds from $J/\psi \to \gamma \pi^+\pi^- \eta'$  decay~\cite{gammacon}. After this, the remaining events for the $J/\psi \to \gamma X(1835)$, $\gamma X(2120)$, and $\gamma X(2370)$ decays are predicted to be $32 \pm 5$, $6 \pm 4$, and $7 \pm 8$, respectively, using the simulated MC samples of corresponding decay processes. The resonant strucures  $f_1(1510)$,  $X(1835)$, $X(2120)$, $X(2370)$ and $X(2600)$ are observed in the $\pi^+\pi^-\eta'$ invariant mass spectra of both the $\eta'$ decay modes of $\eta' \to \gamma \pi^+\pi^-$ and $\pi^+\pi^- \eta$, as seen in Fig.~\ref{dalitz}. The decays of  $J/\psi \to e^+e^- X(1835)$,  $J/\psi \to e^+e^- X(2120)$ and $J/\psi \to e^+e^- X(2370)$ have been observed with significances of $15 \sigma$, $5.3\sigma$ and $7.3 \sigma$, respectively, for the first time by performing a simultaneous fit to the $\pi^+\pi^-\eta'$ invariant mass spectra of both $\eta'$ decay modes and subtracting the gamma conversion events from the corresponding decay processes. The di-electron invariant mass dependent TFF measured in  $J/\psi \to e^+e^- X(1835)$ decays is also reported for the first time. The measured $J/\psi \to e^+e^- X(1835)$ branching fraction is consistent with its theoretical prediction~\cite{JLzhang}. The confirmation of the existence of these resonances in $J/\psi \to e^+e^- \pi^+\pi^- \eta'$  complements the measurements performed in $J/\psi \to \gamma \pi^+\pi^- \eta'$ decays and will be helpful to understand the nature of the $X(1835)$.

\subsection{Search for a light $CP$-odd Higgs boson}
\begin{figure}[htbp]
\centerline{\includegraphics[width=6.0cm]{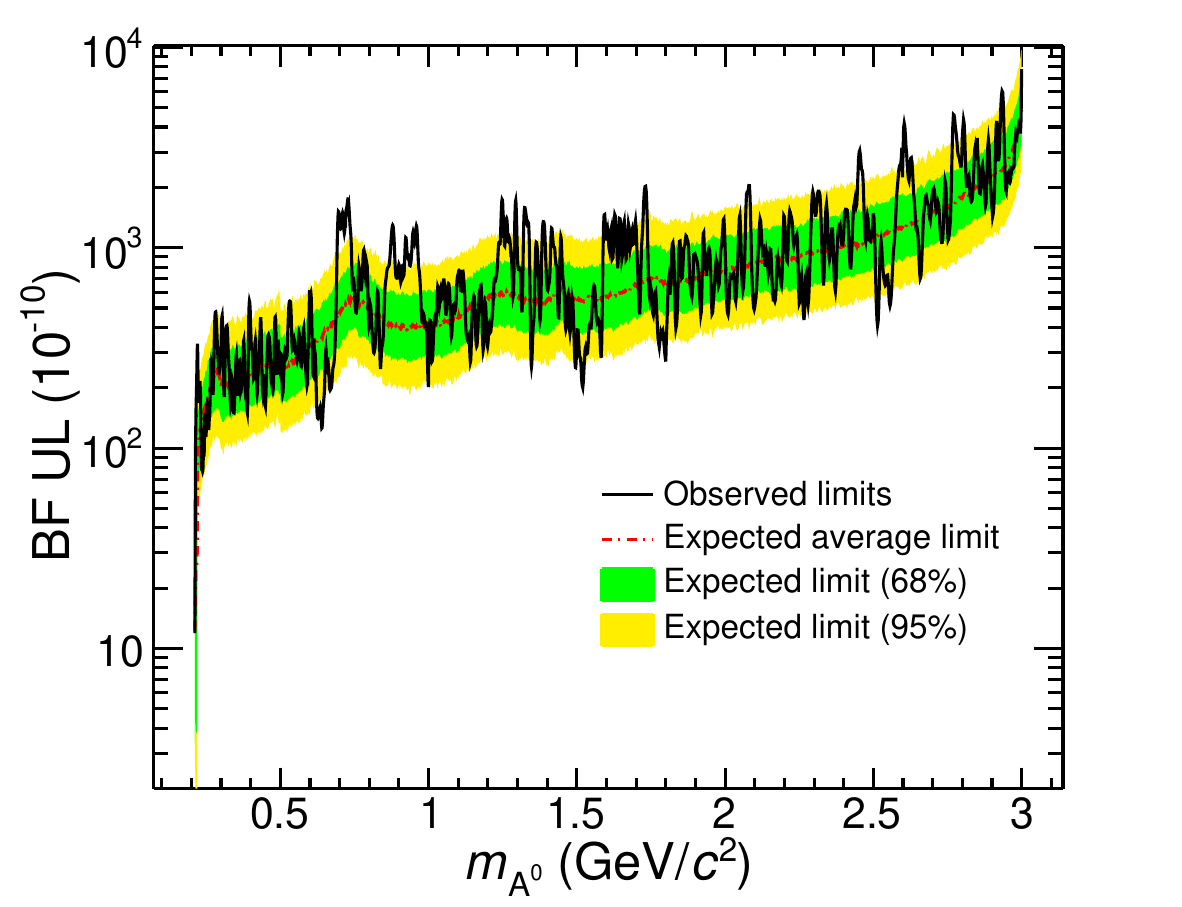}
                \includegraphics[width=6.0cm]{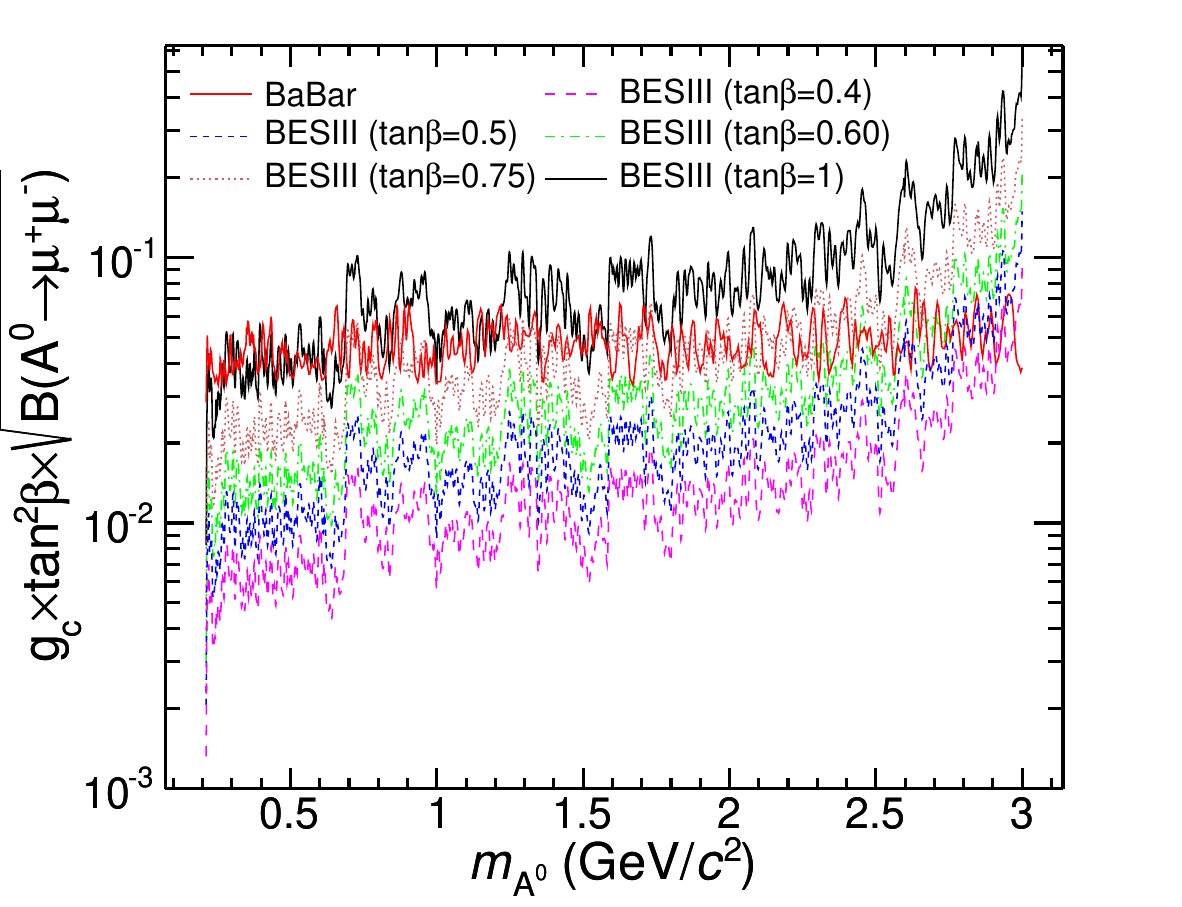}}

\caption{Upper limits on (left)  the product branching fractions $\mathcal{B}(J/\psi \to \gamma A^0) \times \mathcal{B}(A^0 \to \mu^+\mu^-)$ together with expected limit bands at $\pm 1 \sigma$ and $\pm 2 \sigma$ levels, and (right) the effective Yukawa coupling of Higgs field to the bottom quark-pair for different values of $\tan \beta$ together with the  BaBar measurement at $90\%$ CL.  \label{higgs}}
\end{figure}

A light Higgs boson, predicted in many extensions of the SM including NMSSM~\cite{NMSSM1, NMSSM2, NMSSM3}, can be accessible via radiative decays of $J/\psi$.  The coupling of the Higgs field to the up (down) type of quark-pair is proportional to $\cot\beta$ ($\tan \beta$), where $\tan \beta$ is the ratio of the vacuum expectation values of up- and down-type of the Higgs doublets.  The expected branching fraction of $J/\psi \to \gamma A^0$ ranges from $10^{-9}$ to $10^{-7}$, depending upon $m_{A^0}$, $\tan\beta$ and NMSSM parameters~\cite{NMSSM2}. The search for a light Higgs boson has been performed through various decay channels by many collider experiments, including BaBar~\cite{baber} and BESIII~\cite{prevbes3}, only null results have been reported so far.  More recently, BESIII has performed the search for di-muon decays of a light Higgs boson in radiative decays of $J/\psi$ using 9 billion $J/\psi$ decays~\cite{bes3higgs}.  No evidence of significant signal events for  $A^0$ production has been found and set $90\%$ confidence level (CL) upper limits on product branching fractions $\mathcal{B}(J/\psi \to \gamma A^0)\times \mathcal{B}(A^0 \to \mu^+\mu^-)$ have been set in the range $(1.2 -778.0) \times 10^{-9}$ for $0.212 \le m_{A^0} \le 3.0$ GeV/$c^2$.  These limits improve by a factor $6-7$ upon the previous BESIII measurement~\cite{prevbes3} (Fig.~\ref{higgs} (left)) and are slightly more sensitive than the BaBar measurement~\cite{baber} in the low-mass region for $\tan \beta =1$ (Fig.~\ref{higgs} (right)).

\begin{figure}[htbp]
\centerline{\includegraphics[width=12.0cm]{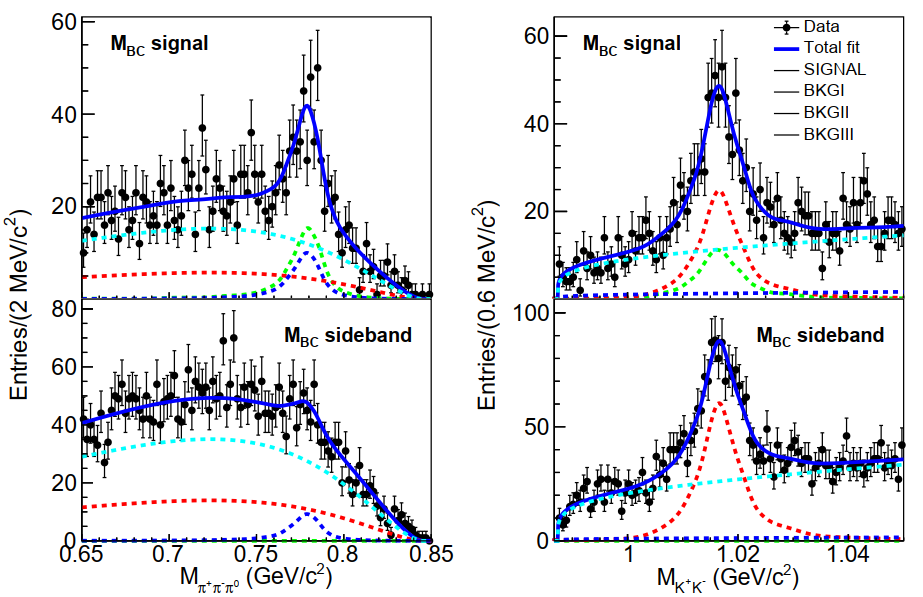}}

\caption{The invariant mass spectra of $\pi^+\pi^-\pi^0$ (left) and $K^+K^-$ (right) systems. The top plots are for the $M_{\rm BC}$ signal region, and the bottom plots are for the $M_{\rm BC}$ sideband region. The black dots with error bars represent the data, the solid blue line represents the total fit, the dotted green curve represents the signal and the dotted blue, the long dotted cyan and the dotted red represent the various background components.    \label{2Dfit}}
\end{figure}

\begin{figure}[htbp]
\centerline{\includegraphics[width=12.0cm]{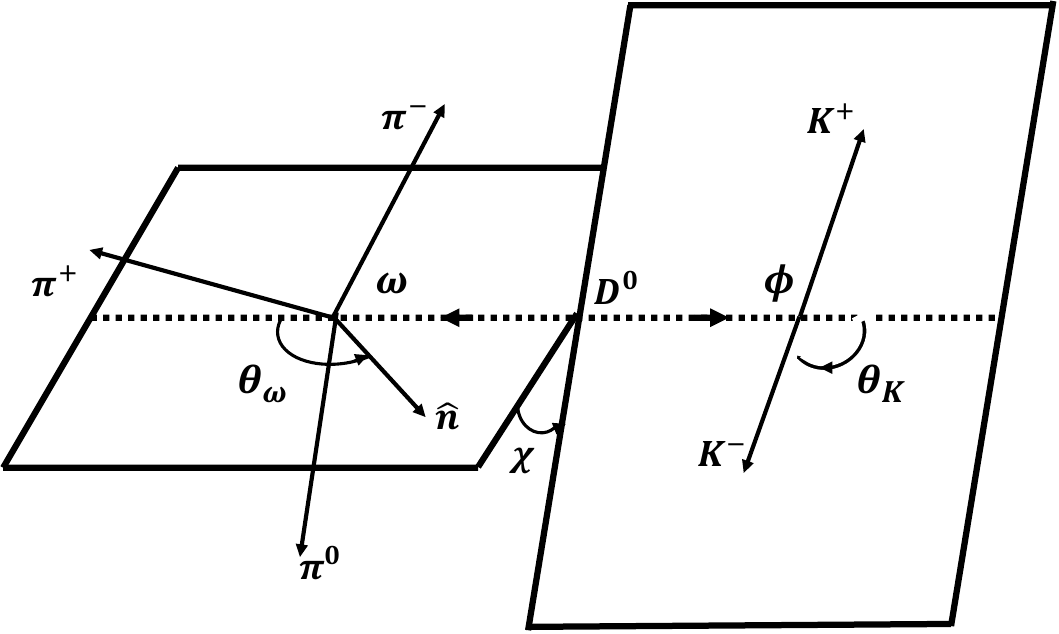}}
\centerline{\includegraphics[width=12.0cm]{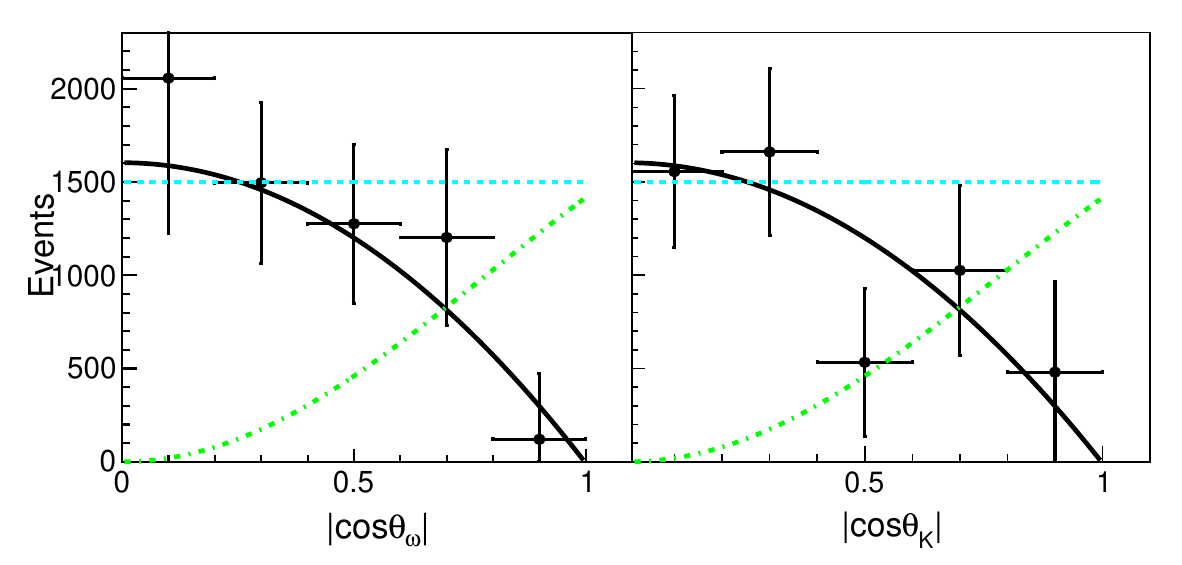}}

\caption{(Top) The $D^0 \to \omega \phi$ decay topology and decay angle definitions. (Bottom) Distributions of efficiency corrected signal events versus $|\cos\theta_{\omega}|$ (left) and $|\cos\theta_K|$ (right). The black dots with error bars represent the data, which includes both statistical and systematic uncertainties, and the solid black curve represents the fit results. The dotted dashed green and dashed cyan curves correspond to the distributions with longitudinal polarization and phase-space assumptions, respectively.  \label{pol}}
\end{figure}

\section{Observation of \boldmath{$D^0 \to \omega \phi$}}
 The measurement of  $D^0 \to \omega \phi$ decays has been performed with 2.93 fb$^{-1}$ of $\psi(3770) \to D^0 \bar{D}^0$ data collected by the BESIII detector using a single tag (ST) technique~\cite{dtowphi}, where $\omega$ and $\phi$ decay to $\pi^+\pi^-\pi^0$ and $K^+K^-$, respectively. One of the $D^0$ mesons is reconstructed with the signal mode of interest and the other $\bar{D}^0$ is allowed to decay inclusively.  The $D^0$ signal is identified by calculating the energy difference $\Delta E = E_D - E_{\rm beam}$ and the beam-constrained mass $M_{\rm BC} = \sqrt{E_{\rm beam}^2/c^4 -p_D^2/c^2}$, where $E_{\rm beam}$ is the beam energy and $E_{D}$ ($p_D$) is the reconstructed energy (momentum) of the $D^0$ candidate in the $e^+e^-$ center-of-mass system.  The $\Delta E$ and $M_{\rm BC}$ distributions peak at zero and the $D^0$ mass position, respectively, for signal-like events. The $D^0 \to \omega \phi$ signal yield has been extracted with a significance of $6.3 \sigma$  by performing a two-dimensional unbinned maximum likelihood fit to the invariant mass spectra distributions of the $\pi^+\pi^-\pi^0$  and $K^+K^-$  systems in both $M_{\rm BC}$ signal and sideband regions as shown in Fig.~\ref{2Dfit}.  The $D^0 \to \omega \phi$ branching fraction has been measured to be $(6.48 \pm 0.96 \pm 0.38) \times 10^{-4}$, where the first and second uncertainties are statistical and systematic, respectively. The measured $D^0 \to \omega \phi$ branching fraction is consistent with the factorization model prediction~\cite{fact}. The polarization of $D^0 \to \omega \phi$ is studied by evaluating the efficiency corrected signal yields in five equal bins of $|\cos\theta_{\omega}|$ and $|\cos\theta_K|$  as shown in Fig.~\ref{pol}, where $\theta_{\omega}$ is the angle in the $\omega$ rest frame between the normal to $\omega$ decay plane and direction of the $D^0$ mesons, and $\theta_K$ is the angle between the direction of $D^0$ meson and one of the $K$ mesons in the $\phi$ rest frame.  The decay $D^0 \to \omega \phi$ appears to be completely transversely polarized, which is inconsistent with the naive factorization~\cite{fact} and Lorentz invariant-based symmetry models~\cite{LI}. These results challenge our understanding of the decay dynamics of charmed meson, and may help in the search for physics beyond the SM.

\section{Summary and future prospects} 
BESIII has conducted a series of studies on light hadron spectroscopy and new physics searches, including the measurements of the $R$ value, radiative $J/\psi$ decays, and hadronic decays of charm mesons using the data samples collected at several energy points, including the $J/\psi$ and $\psi(3770)$ resonances. The $R$ value is measured with an accuracy down to the level of $2.6\%$ for $\sqrt{s} < 3.1$ GeV and $3.0\%$ for $\sqrt{s} > 3.1$ GeV~\cite{bes3rvale}. A new resonance, $X(2600)$, has been observed in the $\pi^+\pi^-\eta'$ invariant mass spectrum in $J/\psi \to \gamma \pi^+\pi^- \eta'$ decay~\cite{bes3x2600} and later confirmed in the electromagnetic Dalitz decay of $J/\psi \to e^+e^- \pi^+\pi^- \eta'$~\cite{EMDalitz}. The $D^0 \to \omega \phi$ branching fraction has been observed for the first time~\cite{dtowphi}. The  decay $D^0 \to \omega \phi$ appears to be completely transversely polarized~\cite{dtowphi}, which is inconsistent with the existing theoretical models~\cite{fact, LI} and challenges our understanding of the decay dynamics of charm mesons. More results on $D^0 \to VV$ are expected to be released in the near future with 20 fb$^{-1}$ of BESIII $\psi(3770)$ data.

\section*{Acknowledgments}
Vindhyawasini Prasad acknowledges the partial financial support received from the Agencia Nacional de Investigación y Desarrollo (ANID), Chile, from ANID FONDECYT regular 1230987 Etapa 2023, Chile, and from ANID PIA/APOYO AFB220004, Chile.


\end{document}